# Nanomechanical sensing using spins in diamond

*Michael S.J. Barson[1†], Phani Peddibhotla[2†], Preeti Ovartchaiyapong[3], Kumar Ganesan[4], Richard L. Taylor[1], Matthew Gebert[1], Zoe Mielens[1], Berndt Koslowski[5], David A. Simpson[4], Liam P. McGuinness[2,4], Jeffrey McCallum[4], Steven Prawer[4], Shinobu Onoda[6], Takeshi Ohshima[6], Ania C. Bleszynski Jayich[3], Fedor Jelezko[2], Neil B. Manson[1] and Marcus W. Doherty[†*]*

1. Laser Physics Centre, Research School of Physics and Engineering, Australian National University, Canberra, ACT 0200, Australia.

2. Institut für Quantenoptik, Universität Ulm, D-89081 Ulm, Germany.

3. Department of Physics, University of California Santa Barbara, Santa Barbara, California 93106, USA.

4. School of Physics, University of Melbourne, Victoria 3010, Australia.

5. Institut für Festkörperphysik, Universität Ulm, D-89081 Ulm, Germany.

6. National Institutes for Quantum and Radiological Science and Technology (QST), 1233 Watanuki, Takasaki, Gunma 370-1292, Japan.

ABSTRACT

Nanomechanical sensors and quantum nanosensors are two rapidly developing technologies that have diverse interdisciplinary applications in biological and chemical analysis and microscopy. For example, nanomechanical sensors based upon nanoelectromechanical systems (NEMS) have demonstrated chip-scale mass spectrometry capable of detecting single macromolecules, such as proteins. Quantum nanosensors based upon electron spins of negatively-charged nitrogen-vacancy (NV) centers in diamond have demonstrated diverse modes of nanometrology, including single molecule magnetic resonance spectroscopy. Here, we report the first step towards combining these two complementary technologies in the form of diamond nanomechanical structures containing NV centers. We establish the principles for nanomechanical sensing using such nano-spin-mechanical sensors (NSMS) and assess their potential for mass spectrometry and force microscopy. We predict that NSMS are able to provide unprecedented AC force images of cellular biomechanics and to, not only detect the mass of a single macromolecule, but also image its distribution. When combined with the other nanometrology modes of the NV center, NSMS potentially offer unparalleled analytical power at the nanoscale.

TEXT

   Nanomechanical sensors based upon nanoelectromechanical systems (NEMS) are a burgeoning nanotechnology with diverse microscopy and analytical applications in biology and chemistry. Two applications with particular promise are force microscopy in geometries that transcend the constraints of conventional atomic force microscopy (AFM)[1-3] and on-chip mass spectrometry with single molecule sensitivity.[4-6] Another burgeoning nanotechnology is quantum nanosensors based upon the electron spin of the NV center in diamond. The NV center has been used to locate single elementary charges,[7] to perform thermometry within living cells[8] and to realize nanoscale MRI in ambient conditions.[9-11] However, there is yet to be a nanosensing application of the NV center that exploits its susceptibility to local mechanical stress/ strain.

Here, we propose that the mechanical susceptibility of the NV center's electron spin can be exploited together with the extreme mechanical properties of diamond nanomechanical structures to realize nano-spin-mechanical sensors (NSMS) that outperform the best available technology. Such NSMS will be capable of both high-sensitivity nanomechanical sensing and the quantum nanosensing of electric fields, magnetic fields and temperature. NSMS will thereby constitute the unification of two burgeoning nanotechnologies and have the potential to perform such unparalleled analytical feats as mass-spectrometry and MRI of single molecules. As the first step towards realizing NSMS, we report the complete characterization of the spin-mechanical interaction of the NV center. We use this to establish the design and operating principles of diamond NSMS and to perform a proof-of-principle demonstration of force measurement using NV centers in diamond microcantilevers. We assess the potential of NSMS for nanomechanical sensing and predict that NSMS are capable of providing unprecedented AC force images of cellular biomechanics and of the inertial imaging of the distribution of mass in single molecules.

The NV center is an atomic-sized defect formed by a nitrogen-substitutional carbon-vacancy pair orientated along the ⟨111⟩ crystallographic direction (see ref 12 for review). Established fabrication methods enable the engineering of NV centers in biocompatible diamond nanostructures.[13] Owing to its bright fluorescence, single NV centers can be optically detected with diffraction-limited resolution. The center's mechanism of optical spin-polarization enables the state of its ground electronic spin triplet to be optically initialized and read out with high fidelity. With the addition of microwave control, quantum metrology techniques may be applied to precisely measure small changes in its spin resonances (see ref 14 for review). This is a method called optically detected magnetic resonance (ODMR). The high sensitivity of NV ODMR is primarily enabled by the center's long-lived spin coherence, which persists in ambient and extreme conditions (up to $T_2^* \sim 200$ μs, $T_2 \sim 3$ ms at room temperature and pressure[15]).

The spin-mechanical interaction of the center was recently investigated in the context of quantum control and computing.[16-21] However, it was incompletely specified, and as a consequence, remains largely uncharacterized (see supplementary information for discussion). Drawing upon the theory of Hughes and Runciman to correctly specify the spin-mechanical interaction,[22] the spin-Hamiltonian of the NV center's ground state electronic spin is

$$H = (D + \mathcal{M}_z)\left(S_z^2 - \frac{2}{3}\right) + \gamma_e \vec{S} \cdot \vec{B} - \mathcal{M}_x(S_x^2 - S_y^2) + \mathcal{M}_y(S_x S_y + S_y S_x), \tag{1}$$

where $D \sim 2.87$ GHz is the zero-field ground state spin-spin interaction, $\vec{S}$ and $\vec{B}$ are the $S = 1$ dimensionless electron spin operators and local magnetic field defined with respect to the NV coordinate system ($xyz$), respectively, $\gamma_e$ is the free electron gyromagnetic ratio, and

$$\begin{aligned}
\mathcal{M}_x &= b(\sigma_{XX} + \sigma_{YY} - 2\sigma_{ZZ}) + c(\sigma_{YZ} + \sigma_{ZX} - 2\sigma_{XY}) \\
\mathcal{M}_y &= \sqrt{3}[b(\sigma_{XX} - \sigma_{YY}) + c(\sigma_{YZ} - \sigma_{ZX})] \\
\mathcal{M}_z &= a1(\sigma_{XX} + \sigma_{YY} + \sigma_{ZZ}) + 2a2(\sigma_{YZ} + \sigma_{ZX} + \sigma_{XY})
\end{aligned} \tag{2}$$

are the components of the spin-mechanical interaction as specified by the stress susceptibility parameters $a1, a2, b$ and $c$, and the components of the local stress tensor $\overleftrightarrow{\sigma}$ defined with respect to the diamond unit cell coordinate system ($XYZ$). See Figure 1a for details of the two coordinate systems. We choose a different coordinate system for the stress tensor to allow simpler relation to laboratory and device design coordinates. Note that the spin-mechanical interaction may be equivalently defined in terms of strain (see supplementary information). For a magnetic field $\vec{B} \approx B_z\hat{z}$, the spin resonances are

$$f_{\pm} = D + \delta \pm \Delta \tag{3}$$

where the shift and splitting of the resonances from their zero-field values are $\delta = \mathcal{M}_z$ and $\Delta = \left[(\gamma_e B_z)^2 + \mathcal{M}_x^2 + \mathcal{M}_y^2\right]^{1/2}$, respectively.

The parameter $a1 = 4.86(2)$ MHz/ GPa has been determined by previous ODMR spectroscopy of NV centers subjected to hydrostatic pressure.[23] To determine the other three susceptibility parameters, we performed ODMR spectroscopy of ensembles of NV centers subjected to uniaxial stress. The ensembles were present in 2×2×2 mm³ cuboid type Ib diamonds with an assortment of {100}, {110} and {111} crystallographic faces. Homogeneous uniaxial stresses with these orientations were generated in the samples by applying pressure to opposing faces using a pneumatic piston-steel anvil arrangement (see Figure 1c and supplementary information). For such stresses, the change in the spin resonances $\Delta f_{\pm}^{app.} = \delta^{app.} \pm \Delta^{app.}$ comprise a shift $\delta^{app.} = \mathcal{M}_z^{app.}$ and a splitting $\Delta^{app.} = \left[\left(\mathcal{M}_x^{app.} + \mathcal{M}_x^{int.}\right)^2 + \left(\mathcal{M}_y^{app.} + \mathcal{M}_y^{int.}\right)^2\right]^{1/2} - \Delta^{int.}$, where $\vec{\mathcal{M}}^{app.}$ are the components that are linear in the applied pressure, $\vec{\mathcal{M}}^{int.}$ are the components that arise from random stress intrinsic to the sample (independent of applied pressure), and $\Delta^{int.} = \left[\left(\mathcal{M}_x^{int.}\right)^2 + \left(\mathcal{M}_y^{int.}\right)^2\right]^{1/2}$ is the intrinsic stress splitting. Figures 1d and 1e show example spectra and the spin resonance shifts and splittings under [110] uniaxial stress. By analyzing the spectra for all three uniaxial stress orientations together, we determined $a2 = -3.7(2), b = -2.3(3),$ and $c = 3.5(3)$ in units of MHz/ GPa (see supplementary information). Given some assumptions, our specification of the spin-mechanical interaction and our parameter values are consistent with the observations of recent investigations (see supplementary information). [18,19]

Having completely characterized the spin-mechanical interaction, we now turn to the design of NSMS. Consider the canonical design of a simple cylindrical pillar with [001] orientation, diameter $w$ and height $h$ (refer to Figure 2a). A transverse force $\vec{F}$ applied to the tip of the pillar bends the pillar, thereby displacing the tip and inducing bending stresses that are maximum at the pillar's base. Applying Euler-Bernoulli beam theory,[24] the displacement of the tip $\vec{r}_{tip}(\vec{F}) = h^3\vec{F}/3EI$ and the bending stress at its base is $\sigma_{ZZ}(\vec{F}) = h\vec{F} \cdot \vec{\xi}/I$, where $E = 1220$ GPa is the Young's modulus of diamond, $I = \pi w^4/64$ is the pillar's moment of area and $\vec{\xi}$ are the internal coordinates of the pillar with respect to ($XYZ$). Assuming the pillar dimensions and the force are sufficiently small that the tip displacement is too small for optical detection, there are two possible approaches to sensing the force via changes in the spin resonances of NV centers: (1) directly measure the spin-mechanical response $\Delta f_{\pm}^{mech} = (a1 \pm 2b)\sigma_{ZZ}(\vec{F})$ of a NV center on the surface of the pillar's base, or (2) applying a transverse magnetic field gradient $\vec{\nabla}B_z$ and measure the spin-magnetic response $\Delta f_{\pm}^{mag} = \pm\gamma_e\vec{\nabla}B_z \cdot \vec{r}_{tip}(\vec{F})$ of a NV center at the pillar's tip. As per our uniaxial stress observations, in the likely case that intrinsic stress dominates $\sigma_{ZZ}(\vec{F})$, the spin-mechanical response reduces to $\Delta f_{\pm}^{mech} \approx a1\sigma_{ZZ}(\vec{F})$ to first order. For both approaches, the change in the spin resonances of the NV centers are linear in $F$ and $r_{tip}$ and their ratio $|\Delta f_{\pm}^{mech}|/|\Delta f_{\pm}^{mag}|$ is plotted in Figure 2b. The plot demonstrates that for nanoscale pillar dimensions ($w, h < 1$ μm), the spin-magnetic response only exceeds the spin-mechanical response if field gradients $|\vec{\nabla}B_z| \gtrsim 30$ mT μm$^{-1}$ are applied. In principle, such field gradients can be applied by positioning nano-ferromagnetic structures within ~100 nm of the nanopillar tip.[10] However, it is known that the thermomagnetic noise and positioning instability of such structures can severely reduce the coherence of NV centers, and thus lead to an overall lower force sensitivity.[25] Furthermore, the need to position the magnetic structure so close to the nanopillar will drastically restrict the range of possible force microscopy geometries. Thus, we conclude that the direct detection of the spin-mechanical response is the more promising approach.

To confirm this conclusion, our characterization of the spin-mechanical interaction and the validity of our simple mechanical model of diamond structures, we performed ODMR spectroscopy of single NV centers in diamond microcantilevers. Forces were applied to the tips of microcantilevers using a motorized tungsten tip (see Figure 2c). Figure 2d shows the measured bending profiles of one of the microcantilevers under different forces obtained via confocal microscopy as well as fits generated by Euler-Bernoulli theory. The fits clearly demonstrate that Euler-Bernoulli theory accurately describes the elastic response of the microcantilever. This accuracy allowed us to extract the applied force directly from the fits $F_{E-B}$. We also used the spin-mechanical response of the NV center in the microcantilever to independently determine the force $F_{ODMR}$. This was done by using our characterization of the NV center's spin-mechanical interaction and application of Euler-Bernoulli theory to determine the stress at the NV center via knowledge of the center's internal position within, the dimensions of and crystallographic orientation of the cantilever. After an ODMR spectrum integration time of 12 minutes, a force uncertainty $\Delta F_{ODMR} \sim 29$ nN was achieved on average. Figure 2e compares $F_{ODMR}$ and $F_{E-B}$ and their agreement clearly confirms the validity of our method.

Given this confirmation, we may now confidently estimate the DC (AC) force sensitivity $\eta_{DC(AC)}$ using the standard photon shot-noise limit expression[26]

$$\eta_{DC(AC)} = \left(2\pi K \frac{d\Delta f_{\pm}^{mech}}{dF}\sqrt{T_{DC(AC)}}\right)^{-1} \quad (4)$$

where $K \approx 0.01$ is the typical value of the factor that accounts for the finite ODMR contrast and photon count rate, and $T_{DC(AC)}$ is the time it takes to perform each measurement shot of the relevant quantum sensing technique. The simplest DC and AC techniques are Ramsey and Hahn-echo, respectively, where $T_{DC} \leq T_2^* \approx 10$ μs and $T_{AC} \leq T_2 \approx 100$ μs are realistic values for NV centers in nanopillars.[18,27] Note that these values for $T_2^*$ and $T_2$ account for the thermomechanical noise of the nanopillar. The smallest shot period is limited to ~10 ns by the shortest practical microwave pulse length, thereby constraining the frequency band for AC force sensing to $\frac{1}{T_{AC}^{max}} \leftrightarrow \frac{1}{T_{AC}^{min}} \approx 10$ kHz $\leftrightarrow$ 100 MHz. Choosing the optimum nanopillar dimensions ($w = 0.1$ μm, $h = 1$μm) that can be fabricated using existing techniques[27] and the optimum value for $T_{DC(AC)}$, then $\frac{d\Delta f_{\pm}^{mech}}{dF} \approx 50$ MHz/ μN and $\eta_{DC(AC)} \approx 100$ pN Hz$^{-1/2}$ (30 pN Hz$^{-1/2}$). These sensitivities define the minimum force $F_{DC(AC)}^{min} = \eta_{DC(AC)}/\sqrt{T_m}$ that can be detected after a measurement averaging time $T_m$. Hence, the nanopillar sensor achieves $F_{DC(AC)}^{min} = 100$ pN within ~1 s (0.1 s) of measurement. Accounting for the differences in geometry and measurement method, our theoretical DC sensitivity estimate for a nanopillar agrees well with the value $\eta_{DC} \sim 40$ nN Hz$^{-1/2}$ we measured in our microcantilever experiments (see supplementary information). Better sensitivities and larger AC frequency measurement bandwidths can be achieved by using more advanced quantum sensing techniques, like D-Ramsey[28] and CPMG,[14] that effectively increase the maximum $T_{DC}$ and $T_{AC}$ beyond the limits imposed by the Ramsey and Hahn-echo techniques. Yet further improvements may be possible using emerging spin-to-charge[29] and electrical[30] spin readout techniques.

We propose that the most promising applications of NSMS in force microscopy are found in cellular biomechanics. Figure 3a compares the size and sensitivity of a NSMS nanopillar sensor (as considered above) to existing techniques for the point-wise measurement of forces in biological systems: conventional AFM, and magnetic and optical tweezing of nanoparticle probes.[34-36] It is clear that, whilst NSMS are possibly smaller probes, they do not necessarily offer superior sensitivity. The advantages of NSMS are rather that (1) there is substantial flexibility in their design owing to the diversity of nanomechanical structures that can be fabricated in diamond[37] and the fact that the probe (ie. tip of nanopillar) is spatially separated from the sensor (ie. NV center at base), and (2) they can be assembled and simultaneously operated *en masse*. These advantages enable NSMS to be engineered for specific microscopy tasks and with extensive multiplexing. An example that demonstrates both of these advantages is an on-chip array of diamond nanopillars for force imaging (depicted in Figure 3b). The nanopillar spacing ($s \gtrsim 250$ nm) is such that quantum sensing can be performed on the single NV centers in each simultaneously using diffraction-limited wide-field optical detection and microwave control. Nanopillar arrays of this kind could be fabricated by combining existing diamond etching[27] and high resolution NV creation[31-33] techniques. Each nanopillar constitutes a force measurement pixel on a wide-field force image. Specifically, the detected change in the spin resonance $\Delta f_{ij} \approx a1h\vec{F}_{ij} \cdot \vec{\xi}_{ij}/I$ of the NV center in the nanopillar at array location $(i,j)$ provides a measure of the projection $\vec{F}_{ij} \cdot \vec{\xi}_{ij}$ of the local force $\vec{F}_{ij}$ onto the position of the center $\vec{\xi}_{ij}$ inside the pillar. Consequently, if the NV center locations are determined by prior characterization, then an image of the force projections can be recovered from the spin resonance measurements with a spatial resolution defined by the nanopillar spacing (ie. ≳250 nm ). Furthermore, if some *a priori* knowledge of the applied force is available, such as it is likely to be constant over the dimensions of four neighboring nanopillars, then the measurements $\overrightarrow{\Delta f_{IJ}} = (\Delta f_{2I-1\ 2J-1}, \Delta f_{2I\ 2J-1}, \Delta f_{2I-1\ 2J}, \Delta f_{2I\ 2J})^{\mathrm{T}}$ of the four neighboring pixels can be combined as a superpixel estimate of the in-plane force vector via $\vec{F}_{IJ} = (a1h/I)^{-1}(\Xi_{IJ}^{\mathrm{T}}\Xi_{IJ})^{-1}\Xi_{IJ}^{\mathrm{T}}\overrightarrow{\Delta f_{IJ}}$, where $\Xi_{IJ} = (\vec{\xi}_{2I-1\ 2J-1}, \vec{\xi}_{2I\ 2J-1}, \vec{\xi}_{2I-1\ 2J}, \vec{\xi}_{2I\ 2J})^{\mathrm{T}}$ and $(I, J)$ are the superpixel indexes. The resulting force vector image will have a spatial resolution defined by the size of the superpixels (ie. ≳500 nm), and using our previous estimates, will achieve a force resolution of 100 pN after ~1 s of measurement time.

The capability of the NSMS nanopillar array to image both DC and AC forces offers an unprecedented means to measure force fields in cellular biomechanics. Existing techniques involve cells being placed upon either an array of compliant nano/micropillars[35] or smooth elastic substrate containing fluorescent markers (ie. Traction Force Microscopy).[36] The cells adhere to the pillars or substrate and the deflection of the pillars or displacement of the substrate fluorescent markers in response to cellular forces are optically tracked. Figure 3a shows that the spatial resolution and sensitivity of the NSMS nanopillar array is comparable to these existing techniques. However, the existing techniques are limited in their measurement of AC forces because they require sub-diffraction limit optical resolution of fast deflections/ displacements, which is constrained by the brightness and fluctuations of the fluorescent markers and background. Indeed, to our knowledge, AC forces are yet to be imaged. The NSMS nanopillar array has a marked advantage because it does not suffer from this limitation since AC forces are instead detected by the spin of the embedded NV center. The imaging of AC forces by NSMS nanopillar arrays will allow time-resolved spatial correlation of the dynamics of the cells, which may provide new insight into the mechanisms driving such phenomena as cell migration. Yet further insight is likely to be gained by also incorporating NV centers into the tips of the nanopillars and using these NV centers to concurrently perform nano-thermometry/-electrometry/-magnetometry of the cell.

We propose that NSMS constructed from diamond nanobeams with embedded NV centers cannot only achieve inertial detection of single macromolecules for mass spectrometry, but also realize the inertial imaging of the distribution of mass in a macromolecule for enhanced molecular identification and analysis. Drawing upon the theory of NEMS inertial detection/imaging,[6] to first order, a molecular adsorbate changes the frequency $v_n$ and displacement profile $\psi_n(x)$ of the $n^{th}$ out-of-plane flexural vibrational mode of a nanobeam by

$$\Delta v_n \approx -v_n \int_0^1 \frac{\mu_1(x)}{m_0} \psi_n^2(x) dx$$
$$\Delta \psi_n(x) \approx \sum_{m \neq n} c_{nm} \psi_m(x) = \sum_{m \neq n} \left[ \frac{v_n^2}{v_m^2 - v_n^2} \int_0^1 \frac{\mu_1(x)}{m_0} \psi_m(x) \psi_n(x) dx \right] \psi_m(x) \quad (5)$$

where $l$ is the length of the nanobeam, $m_0$ is the nanobeam's (assumed uniformly distributed) mass, $\mu_1(x)$ is the adsorbate mass distribution along the nanobeam, $c_{nm}$ is the coupling coefficient of the $n^{th}$ and $m^{th}$ modes, and the modes satisfy the orthonormality condition $\int_0^l \psi_n(x) \psi_m(x) dx = \delta_{nm} l$. Existing NEMS methods measure the frequency shift $\Delta v_n$ by detecting the change in the time-dependent amplitude $A_n(t)$ or phase of the $n^{th}$ mode when it is driven by a known on-resonant AC force $F_n(t) = F_n \cos 2\pi v_n t$ with fixed amplitude and frequency.[6] To first order, the vibrational amplitude is

$$A_n(t) = A_n[U \cos 2\pi v_n t + V \sin 2\pi v_n t] \approx 2Q_n A_n \frac{\Delta v_n}{v_n} \cos 2\pi v_n t \quad (6)$$

where $U = \frac{(\Delta v_n/v_n)(1/2Q_n)}{(\Delta v_n/v_n)^2+(1/2Q_n)^2}$ and $V = \frac{(1/2Q_n)^2}{(\Delta v_n/v_n)^2+(1/2Q_n)^2}$ are the in- and out-of-phase vibrational components, $A_n = Q_n F_n/m_0 v_n^2$ is the vibrational amplitude in the absence of the adsorbate and $Q_n$ is the Q-factor of the $n^{th}$ mode. Since these existing methods measure just the frequency shifts $\Delta v_n$ and not the coupling coefficients, they are constrained to estimating the adsorbate's moments of mass[6]

$m^{(k)} = \int_0^l \mu_1(x) x^k dx$ via

$$m^{(k)} = \int_0^l \mu_1(x)\left[\sum_n \alpha_{kn}\psi_n^2(x)\right]dx \approx m_0 \sum_n \alpha_{kn}\frac{\Delta v_n}{v_n} \tag{7}$$

where, using knowledge of $\psi_n(x)$, the linear coefficients $\alpha_{kn}$ are defined such that $x^k \approx \sum_n \alpha_{kn}\psi_n^2(x)$. In this way, the estimate of each mass moment improves with the number of modes whose frequency shifts have been measured. As a consequence, the method is constrained by the bandwidths of mechanical sources available to drive the resonances and the typical reduction of Q-factor with increasing mode frequency.

The NEMS method can be implemented using NSMS nanobeams as well. Excitation of the $n^{th}$ mode of the nanobeam generates a time-varying stress $\sigma_{n,XX}(\vec{\xi},t) = A_n(t)E\xi_z \left.\frac{d^2\psi_n}{dx^2}\right|_{\xi_x} + \sum_{m\neq n} c_{nm}\left.\frac{d^2\psi_m}{dx^2}\right|_{\xi_x}$ at the NV center located at $\vec{\xi}$ (see Figure 4a for geometry).[24] The first-order time-varying changes of the NV center's spin resonances are then

$$\Delta f_\pm^{mech}(t) \approx 2Q_n\Phi_n(\vec{\xi})\frac{\Delta v_n}{v_n}\cos 2\pi v_n t + \sum_{m\neq n} c_{nm}\Phi_m(\vec{\xi})\sin 2\pi v_n t \tag{8}$$

where $\Phi_n(\vec{\xi}) = A_n a 1 E\xi_z \left.\frac{d^2\psi_n}{dx^2}\right|_{\xi_x}$. Thus, the frequency shift $\Delta v_n$ of the $n^{th}$ mode can be measured by performing a repetitive Hahn-echo (or higher order) sequence with the same period as, but out-of-phase with, the on-resonant mechanical driving of the mode (see Figure 4b). The change in the quantum phase accumulated during the Hahn-echo sequence caused by the frequency shift is $\Delta\Theta_n^{out} = 2Q_n\Phi_n(\vec{\xi})\Delta\nu_n/\nu_n$. It follows that the shot-noise limited detection sensitivity of $\Delta\nu_n/\nu_n$ is

$$\eta_{\Delta\nu_n/\nu_n} = \left[2\pi K \frac{d\Delta\Theta_n^{out}}{d\Delta\nu_n/\nu_n}\sqrt{1/\nu_n}\right]^{-1} = \sqrt{\nu_n}\left[4\pi K Q_n\Phi_n(\vec{\xi})\right]^{-1}$$

and thus, the mass sensitivity is $\eta_{mass} = m_0\eta_{\Delta\nu_0/\nu_0} \approx 1$ zg Hz$^{-1/2}$ for a nanobeam of 0.1×0.1×5 µm dimensions, an optimally located NV center (ie. at positions of greatest bending stress: $\xi_x = 0, \frac{l}{2}$, or $l$ and $\xi_z = \pm h/2$), and a typical $Q_0 \sim 100$ in air.[19] This sensitivity corresponds to the detection of a mass equivalent to 50 carbon atoms (ie. smaller than a single protein) within ~1 s, which is comparable to current predictions for NEMS employing carbon nanotubes (see Figure 4c).[6]

Unlike existing NEMS techniques, NSMS can also measure the mode coupling coefficients $c_{nm}$.[6] This is achieved by performing the Hahn echo sequence in-phase with the mechanical driving, rather than out-of-phase as before. The change in the Hahn-echo accumulated phase caused by the coupling of the $n^{th}$ mode to other modes is $\Delta\Theta_n^{in} = \sum_{m\neq n} c_{nm}\Phi_m(\vec{\xi})$. Thus, given prior knowledge of $\Phi_m$ and by measuring the ODMR change for many vibrational modes, the coupling coefficients may be estimated. Alternatively, by applying a magnetic field gradient $\partial_x B_z \gtrsim 0.1$ mT µm$^{-1}$ to spectrally distinguish NV centers along the length of the nanobeam and detecting the change in the ODMR signal of each $\Delta\Theta_{n,i}^{in} = \sum_{m\neq n} c_{nm}\Phi_m(\vec{\xi_i})$ (see Figure 4d), then fewer vibrational modes need to be interrogated to estimate the same number of coefficients, which has distinct advantages as discussed previously. The application of such magnetic field gradients to NV centers has been demonstrated using microcoils.[11] The unique capability of NSMS to determine the mode frequency shifts as well as their coupling coefficients allows direct imaging of the adsorbate mass distribution (rather than evaluating its mass moments) via

$$\mu_1(x) = \int_0^l \mu_1(x')\delta(x-x')dx' \approx -m_0\left[\sum_n \beta_n^2(x')\frac{\Delta v_n}{v_n} + \sum_{n,m}\beta_n(x')\beta_m(x')\left(\frac{v_n^2-v_m^2}{v_n^2}\right)c_{nm}\right] \quad (9)$$

where the linear coefficients $\beta_n(x')$ are chosen such that $\delta(x-x') \approx \left[\sum_n \beta_n(x')\psi_n(x)\right]^2$. The spatial resolution of this mass imaging is ultimately defined by the fidelity of this delta function approximation achieved for a given number of measured coupling coefficients. For measurements of the first $N$ modes, the delta function is approximated by a Gaussian of width $\alpha \approx l/2\pi N$ and so, resolution comparable to molecular sub-structures $\alpha < 10$ nm can be achieved for $N > 10$ and $l = 5$ μm.

To demonstrate the analytical power possible by combining NSMS inertial imaging with the other nanometrology applications of the NV center, consider the magnetic resonance imaging of the inertially detected molecular adsorbate. This would simply require the mechanical driving to be turned off to stabilize the distance between the adsorbate and the NV centers in the nanobeam. The magnetic field gradient would be retained as per ref 11. Similarly, NV electrometry could instead be employed to identify distinct electrostatic features of the molecular adsorbate, such as active sites of proteins. Indeed, there are likely to be many more powerful combinations of sensing modalities that employ more sophisticated nanomechanical architectures than the simplest ones considered here. There are clearly burgeoning opportunities in a new field of NSMS.

FIGURES

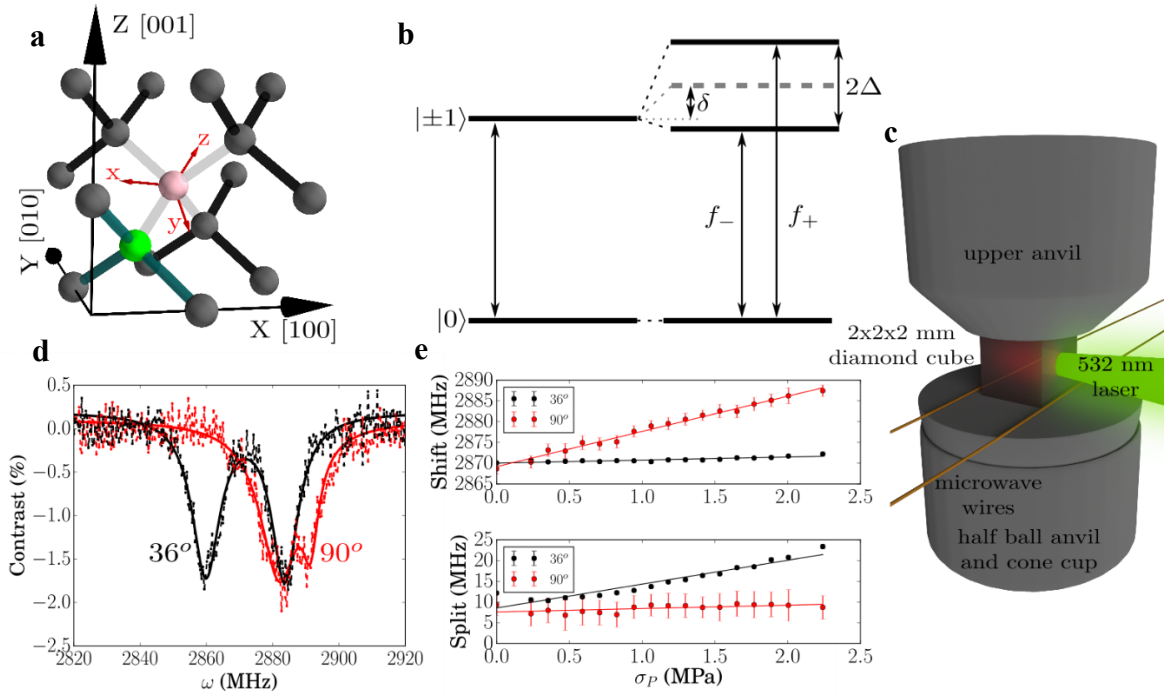

**Figure 1.** The spin-mechanical interaction of the NV center. (a) diagram of the NV center in the diamond unit cell denoting the coordinate systems of the NV center (*xyz*) and the unit cell (*XYZ*). (b) ground state fine structure diagram denoting the spin resonances $f_\pm$ and the mechanically induced shift $\delta$ and splitting $\Delta$. (c) schematic of the uniaxial stress experiment. (d) and (e), example spectra and spin resonance shifts and splittings under [110] uniaxial stress. The two distinct spectra and sets of shifts and splittings correspond to those NV centers whose orientations are inclined by 90° and 36° from [110].

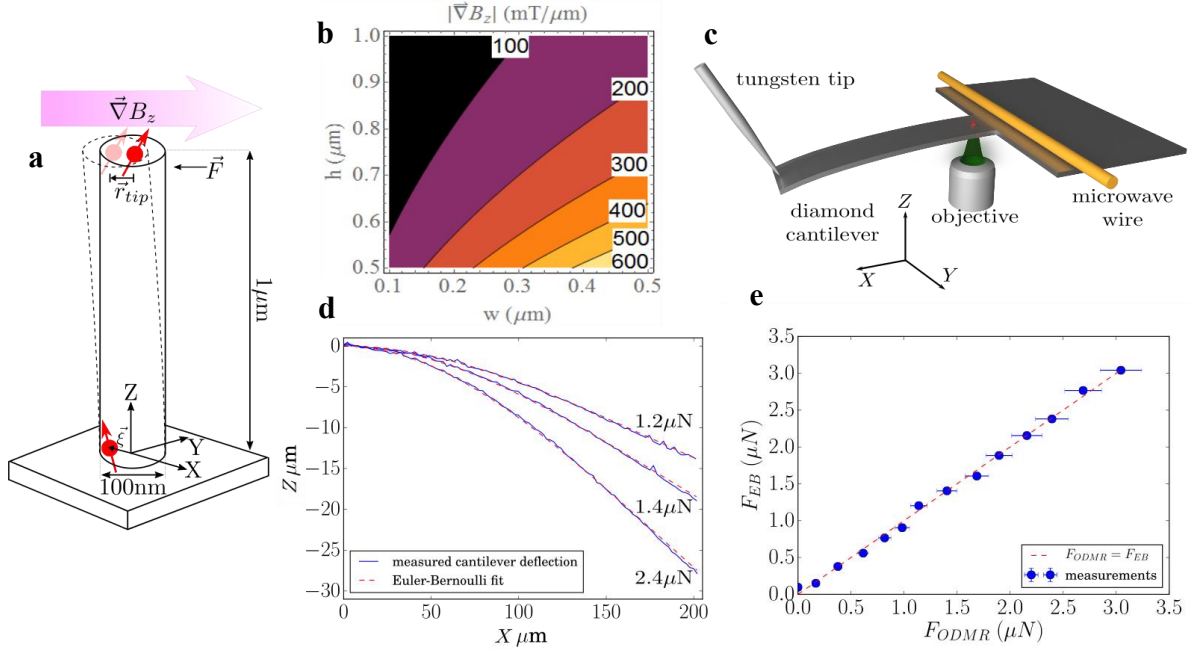

**Figure 2.** The diamond nanopillar spin-mechanical sensor. (a) diagram of the canonical design of a diamond nanopillar NSMS as described in main text. (b) contour plot of the magnetic field gradient $\vec{\nabla} B_z$ that is required for the spin-magnetic response $\Delta f_{\pm}^{mag}$ of the NV center at the nanopillar's tip to exceed the spin-mechanical response $\Delta f_{\pm}^{mech}$ of the NV center at the nanopillar's base. (c) schematic of the microcantilever experiment. Note that the cantilever coordinate system ($XYZ$) is defined such that $X \parallel [110]$ and $Z \parallel [001]$. (d) optical measurements of the bending of the microcantilever under different forces applied to its tip, including fits obtained using Euler-Bernoulli theory. (e) comparison of the force as inferred from the Euler-Bernoulli fits in (d) with the force as inferred from the combination of ODMR measurements of the embedded NV center and the Euler-Bernoulli model of the microcantilever's mechanics.

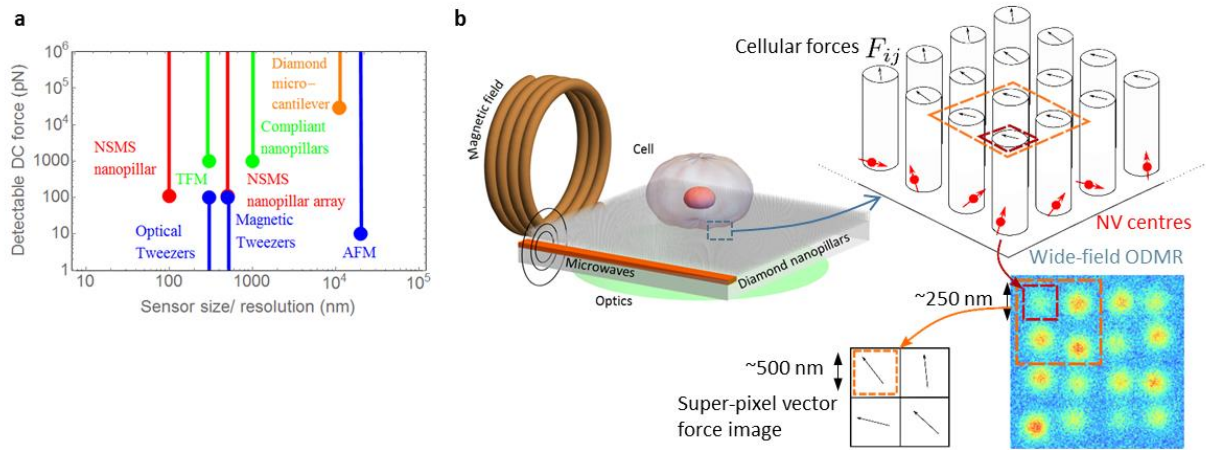

**Figure 3.** Nano-spin-mechanical sensors for bio-force microscopy. (a) comparison of the size/ spatial resolution (points) and range (indicated by connecting lines) of detectable DC forces of individual NSMS nanopillars (red)/ nanopillar arrays (red), the diamond microcantilever measured in this work (orange) and existing techniques (blue) for point-wise force sensing/ imaging (green). The values for existing techniques were collected from refs 34-36. (b) schematic of the NSMS nanopillar array for imaging the forces exerted by living cells as described in the text.

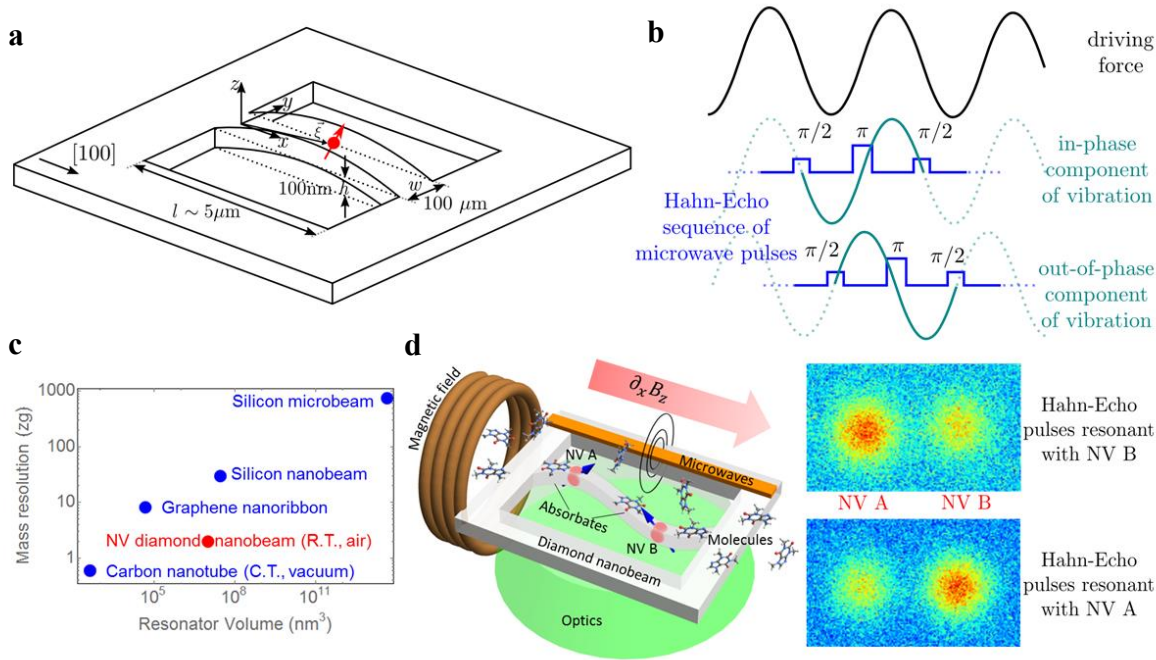

**Figure 4.** Nano-spin-mechanical sensors for inertial imaging of molecules. (a) diagram of a diamond nanobeam NSMS for the inertial imaging of molecules as described in main text. (b) schematic of the phase relationships between the AC driving force, the in- and out-of-phase components of the nanobeam's vibrational response, and the in- and out-of-phase Hahn-echo microwave pulse sequences used to selectively measure the components' amplitudes via the embedded NV center's ODMR. (c) comparison of the resonator volume and mass resolution of the NSMS nanobeam at room temperature in air (red) (after 1s of averaging) and NEMS (blue) including the projected resolution of carbon nanotubes at cryogenic temperatures in vacuum.[6] (d) schematic of the enhanced inertial imaging technique that uses multiple NV centers, a magnetic field gradient and microwave frequency selection to determine the mode coupling coefficients. The right panel demonstrates how microwave frequency selection enables ODMR measurements to be performed on individual NV centers in the nanobeam.

## ASSOCIATED CONTENT

**Supporting Information**

Additional information on experimental methods and analysis, and theoretical models and estimations.

## AUTHOR INFORMATION

**Corresponding Author**

*E-mail: marcus.doherty@anu.edu.au.

**Author Contributions**

The manuscript was written through contributions of all authors. All authors have given approval to the final version of the manuscript. †These authors contributed equally.

**Notes**

The authors declare no competing financial interests.

## ACKNOWLEDGMENT

We acknowledge support by the ARC (DP140103862), the DAAD-Go8 Cooperation Scheme, the Air Force Office of Scientific Research MURI programme, DFG (SFB/TR21, FOR1493), Volkswagenstiftung, EU (DIADEMS, SIQS) and ERC. We thank P. Maletinsky for the provision of a cantilever sample (although not used in final experiments). We also thank P. Maletinsky, J. Wrachtrup, M.J. Sellars, S.A. Momenzadeh and Z. Chu for fruitful discussions.

[16] Bennett, S. D.; Yao, N. Y.; Otterbach, J.; Zoller, P.; Rabl, P.; Lukin, M. D. Phonon-induced spin-spin interactions in diamond nanostructures: Application to spin squeezing. *Phys. Rev. Lett.* **2013,** *110*, 156402.

[17] MacQuarrie, E. R.; Gosavi, T. A.; Jungwirth, N. R.; Bhave, S. A.; Fuchs, G. D. Mechanical spin control of nitrogen-vacancy centers in diamond. *Phys. Rev. Lett.* **2013,** *111*, 227602.

[18] Ovartchaiyapong, P.; Lee, K. W.; Myers, B. A; Bleszynski Jayich, A. C. Dynamic strain-mediated coupling of a single diamond spin to a mechanical resonator. *Nature Comm.* **2014,** *5*, 4429.

[19] Teissier, J.; Barfuss, A.; Appel, P.; Neu, E.; Nunnenkamp, A.; Maletinsky, P. Strain coupling of a nitrogen-vacancy center spin to a diamond mechanical oscillator. *Phys. Rev. Lett.* **2014,** *113*, 020503.

[20] Barfuss, A.; Teissier, J.; Neu, E.; Nunnenkamp, A.; Maletinsky, P. Strong mechanical driving of a single electron spin. *Nature Phys.* **2015,** *11*, 820.

[21] Golter, D. A.; Oo, T.; Amezcua, M.; Stewart, K. A.; Wang, H. Optomechanical quantum control of a nitrogen-vacancy center in diamond. *Phys. Rev. Lett.* **2016,** *116*, 143602.

[22] Hughes, A. E.; Runciman, W. A. Uniaxial stress splitting of doubly degenerate states of tetragonal and trigonal centres in cubic crystals. *Proc. Phys. Soc.* **1967,** *90*, 827.

[23] Doherty, M. W.; Struzhkin, V. V.; Simpson, D. A.; McGuinness, L. P.; Meng, Y.; Stacey, A.; Karle, T. J.; Hemley, R. J.; Manson, N. B.; Hollenberg, L. C. L.; Prawer, S. Electronic
22

TABLE OF CONTENTS GRAPHIC

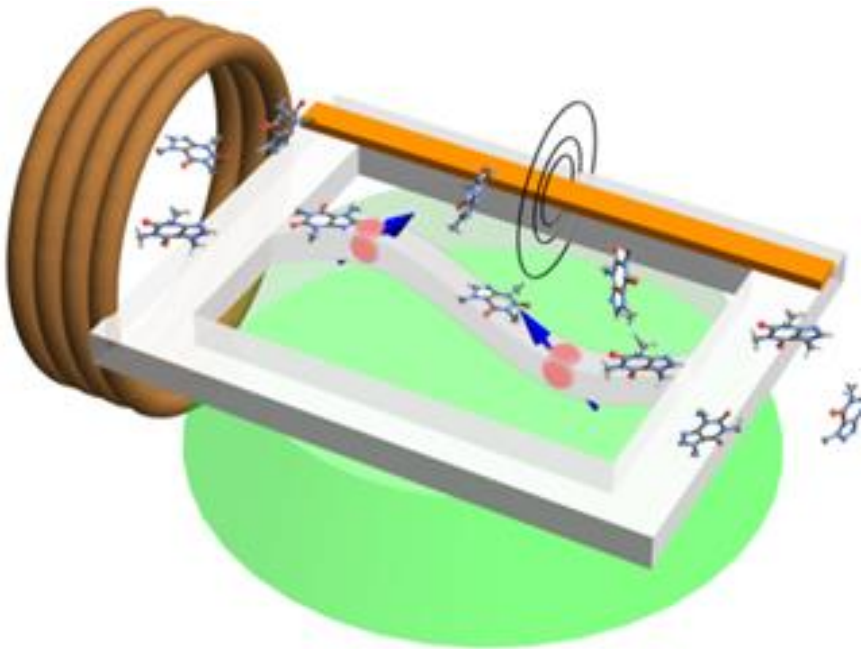